# DENSITY SCALING AND DYNAMIC CORRELATIONS IN VISCOUS LIQUIDS

D. Fragiadakis, R. Casalini, and C.M. Roland

*Naval Research Laboratory, Code 6120, Washington DC 20375-5342*

**Abstract**

We use a recently proposed method [Berthier L.; Biroli G.; Bouchaud J.P.; Cipelletti L.; El Masri D.; L'Hote D.; Ladieu F.; Pierno M. *Science* **2005**, *310*, 1797.] to obtain an approximation to the 4-point dynamic correlation function from derivatives of the linear dielectric response function. For four liquids over a range of pressures, we find that the number of dynamically correlated molecules, $N_c$, depends only on the magnitude of the relaxation time, $\tau_\alpha$, independently of temperature and pressure. This result is consistent with the invariance of the shape of the relaxation dispersion at constant $\tau_\alpha$ and the density scaling property of the relaxation times, and implies that $N_c$ also conforms to the same scaling behavior. For propylene carbonate and salol $N_c$ becomes constant with approach to the Arrhenius regime, consistent with the value of unity expected for intermolecularly non-cooperative relaxation.

**Introduction**

A liquid's approach to vitrification is associated with various property changes, the most significant being the enormous increase in the reorientational (or structural) relaxation time, $\tau_\alpha$. The glass transition is commonly defined as the condition of temperature, $T$, and pressure, $P$, at which $\tau_\alpha$ assumes an arbitrarily long value, e.g., 100 s; however, the importance of $\tau_\alpha$ extends beyond serving as a signature of $T_g$. The Kohlrausch exponent, $\beta$, describing the breadth of the relaxation function, is a unique function of $\tau_\alpha$ [1,2], as is the thermodynamic quantity $TV^\gamma$, where $V$ is the specific volume and $\gamma$ a material constant [3,4,5]. Thus, the distribution of relaxation times (reflected in $\beta$) and the set of state points associated with a given distribution are both determined entirely by the magnitude of $\tau_\alpha$. Additionally, different ranges of temperature and pressure are associated with different regimes of dynamical behavior: At $T_A$ the temperature dependence of $\tau_\alpha$ becomes non-Arrhenius [6,7] and at $T_B$ the "dynamic crossover" is observed [8]. An intriguing feature of these dynamic transitions is that the relaxation time at both is a material constant; changes in pressure or volume change $T_A$ and $T_B$, but the relaxation times at either characteristic temperature do not change [9,10,11].

The many-body dynamics of viscous liquids are necessarily intermolecularly correlated and dynamically heterogeneous. Theories of the glass transition often posit a diverging dynamic length scale as the origin of vitrification [12,13,14,15,16,17,18,19], with a concomitant increase in $\tau_\alpha$ as a larger number of molecules must be accommodated for relaxation to occur. This naturally leads to the expectation of a connection between the relaxation times and the dynamic correlation length. While the existence of spatially heterogeneous dynamics is well accepted [20,21,22], experimentally quantifying their size and behavior is difficult because it requires consideration of both spatial and temporal correlations. The spatial extent of the fluctuations over a time span $t$ can be characterized using a four-point susceptibility defined as [23,24,25]

$$\chi_4(t) = \int \langle \rho(r_1,0)\rho(r_1+r_2,0)\rho(r_1,t)\rho(r_1+r_2,t)\rangle_{r_1} dr_2 \qquad (1)$$

$\chi_4(t)$ exhibits a maximum at $t \sim \tau_\alpha$ that is proportional to the number of molecules, $N_c$, dynamically correlated over this time span [17,26,27].

A recent advance in experimental characterization of the dynamic correlations was the derivation by Berthier and coworkers [24,28] of $\chi_4(t)$ in terms of the temperature derivative of a two-point dynamic correlation function

$$\chi_4(t) \geq \frac{k_B}{c_P}T^2\chi_T^2(t) = \frac{k_B}{c_P}T^2\left(\frac{\partial \Phi(t)}{\partial T}\right)^2 \qquad (2)$$

where $\Phi(t)$ is an experimentally accessible susceptibility. It has been found that at lower temperatures the inequality can be replaced by an equal sign [28,29] and to a good approximation

$$N_c(T,P) \approx \frac{k_B}{\Delta c_P(P)}T^2\left\{\max_t \chi_T(t,T,P)\right\}^2 \qquad (3)$$

In this equation $\Delta c_P$ is the isobaric heat capacity increment at $T_g(P)$ and $\max_t \chi_T$ represents the maximum value of $\chi_T(t)$ for any state point.

In this note we analyze dielectric relaxation spectra to determine $N_c$ as a function of temperature at various pressures. While it has been found that in the limit of zero pressure the spatial extent of the dynamic correlation increases monotonically with the relaxation time [28], we show that for a given material $\tau_\alpha$ uniquely determines $N_c$ for any thermodynamic state. This result is consistent with the fact that $\beta$ is determined by $\tau_\alpha$ [1,2] and the previously deduced scaling law for $\tau_\alpha$ [3,4,5].

**Method**

Normalized dielectric spectra measured over a broad range of temperatures and pressures were fit in the frequency domain to the Fourier transform of the Kohlrausch function $\Phi(t) = \exp[-(t_\alpha/\tau)^\beta]$. Similar to the method used in ref. [28] to determine $N_c(T)$, the temperature and pressure dependences of the parameters $\tau_\alpha$ and $\beta$ were interpolated using a polynomial or other function (e.g., the VFTH [30] and Avramov [13] equations were used for $\tau_\alpha$). From this parameterization we computed $\chi_T(t)$ at various temperatures and pressures by taking the analytical derivative of the parameterized $\Phi(t,T,P)$. The calculations of $\chi_T(t)$ were limited to the range of the experimentally measured data.

The analysis was carried out on four molecular liquids: salol, propylene carbonate (PC), a polychlorinated biphenyl having 54% by weight chlorine (PCB), and a mixture of 67% *o*-terphenyl and 33% *o*-phenylphenol (OTP-OPP). For the OTP-OPP, directly measured heat capacity data at elevated pressure are available [31]. For salol, propylene carbonate, and PCB, $\Delta c_P(P)$ was calculated from the heat capacity at ambient pressure (refs. [32,33,34] respectively) and the equation of state (refs. [35,36,37] respectively) using the Maxwell relation

$$\left(\frac{\partial c_P}{\partial P}\right)_T = -T\left(\frac{\partial^2 V}{\partial T^2}\right)_P \qquad (4)$$

with the assumption of a negligible change with pressure in the heat capacity of the glass at $T_g$. This assumption is supported by the $c_P(P)$ data in ref. [31] for OTP-OPP and polystyrene. Thus,

$$\Delta c_p(P) = \Delta c_p(P=0) - \int_0^P \left(T\frac{\partial^2 V(T)}{\partial T^2}\right)_{T_g} dP' \qquad (5)$$

where the second term on the rhs is for the liquid only. The dielectric spectra were obtained from the literature: OTP-OPP [38], salol [39], PC [36,40], and PCB [37].

**Results**

Figure 1 shows $N_c$ versus temperature at the lowest and highest pressures (values of $P$ given in the figure caption) for each of the four liquids. For a given temperature the differences in $N_c$ for high and low pressure can exceed one decade, indicating the substantial dependence on $P$ over the studied range. When plotted versus $T$ there is no indication of a divergence in the correlation size as temperature is lowered (as recently deduced for $\tau_\alpha$ as well [41]). On the other hand, for PC at higher values of $T$ the number of dynamically correlated molecules becomes less than one, an unphysical result that is discussed below.

In Figure 2 the number of correlated molecules is plotted versus $\tau_\alpha$. For each liquid the relaxation time defines $N_c$; that is, any variation in $(N_c)_{\tau_\alpha}$ with pressure is less than the experimental error (which is estimated to be at about 10 - 20%). The near invariance of $N_c$ at fixed $\tau_\alpha(T,P)$ is consistent with recent molecular dynamic simulations of Lennard-Jones particles showing the maximum in $\chi_4(t)$ to be constant at fixed $\tau_\alpha$ [42]. It is also an expected consequence of the scaling relation [3,4,5]

$$\tau_\alpha = f(TV^\gamma) \tag{6}$$

To see this, from eqs. (2) and (3)

$$N_c \sim T^2 \max\left(\frac{\partial \Phi}{\partial T}\right)^2 \tag{7}$$

and

$$T\frac{\partial \Phi}{\partial T} = \frac{\partial \Phi}{\partial (TV^\gamma)} TV^\gamma (1+\gamma\alpha_P T) \tag{8}$$

where $\alpha_P$ is the isobaric thermal expansion coefficient. Since the (normalized) Φ(t) depends only on $\tau_\alpha$ [1,2] (which itself implies $N_c$ is constant at fixed $\tau_\alpha$ if the nonexponentiality of Φ(t) is attributed to dynamic heterogeneity), it follows from eq. (6) that

$$(N_c)_{\tau_\alpha} \propto (1+\gamma\alpha_P T)^2 \tag{9}$$

Thus, at fixed $\tau_\alpha$ (fixed $TV^\gamma$) $N_c$ varies with $P$ only by the factor $(1+\gamma\alpha_P T)^2$. Since the product of $\alpha_P T$ at $T_g$ is essentially a universal constant (Boyer-Spencer rule [43]), we expect $(N_c)_{\tau_\alpha}$ to be constant at least at $T_g$. At any $\tau_\alpha$ other than $\tau_\alpha(T_g)$, the variation of $N_c$ with $P$ can be assessed from the variation with pressure of the isobaric fragility, since the latter is proportional to $(1+\gamma\alpha_P T)$ [44]. Experimentally it has been found for several liquids, including PC, salol, and a polychlorinated biphenyl, that up through pressures as high as 500 MPa the fragility varies only about 10% (Figure 10 in ref. [44]). This is within the uncertainty in the analyses herein, thus accounting for the invariance of $N_c$ at fixed $\tau_\alpha$ in Fig. 2. To a good approximation the spatial extent of the dynamic correlations follows the same scaling law of $\tau_\alpha$; that is,

$$N_c \cong g(TV^\gamma) \tag{10}$$

although the function $g$ is not the same as the $f$ in eq. (6).

The interest in dynamic correlation arises from the idea that a growing spatial extent of intermolecular cooperativity underlies the supercooled dynamics, perhaps leading ultimately to vitrification [17,18,19,41]. It is also commonly believed that the change in dynamics at $T_B$ is a consequence of a growth of the correlation volume that commenced with onset of non-Arrhenius behavior at $T_A$ [6,7,8,45]. These ideas lead to two expectations: The increase in $N_c$ as $T_g$ is approached should parallel the effect on the Kohlrausch exponent $\beta$, which presumably reflects inversely the extent of the dynamic heterogeneity. And at the opposite end of the range of $\tau_\alpha$, $N_c$ should go to unity at high temperatures.

We examine these expected trends in Figure 3, showing an Arrhenius plot of $N_c$ at $P$ = 0.1 MPa for the three liquids herein for which the experimental $\tau_\alpha$ have been measured above $T_B$. The general trends are in accord with previous results [19,28,46]. For both PCB and salol the rate of growth of $N_c$ decreases on cooling below $T_B$, mirroring the trend of the Kohlrausch exponent $\beta$ characterizing the relaxation time distribution. An increasingly non-Arrhenius behavior is thus associated with a slowing of the growth of dynamic correlations. This ostensible disconnect between dynamic heterogeneity and the distribution of relaxation times has been seen previously in *o*-terphenyl [47], tris(naphthyl)benzene [48,49,50], and sucrose benzoate [51], for which $\beta$ is unchanged with temperature variations over the range of $T/T_g$ from *ca.* 1.22 to 1.02.

We also note that $N_c$ for PC becomes smaller than one, which cannot be. This result has been observed previously and ascribed to *"prefactors* [and] *normalization, etc., which might affect the vertical axis"* [28]. For PCB in Fig. 3 there is a steep slope for $N_c(T)$ above $T_B$, with no suggestion of $N_c$ leveling off at a constant value (equal to unity with proper normalization). However, for PC the data extend above $T_A$ (= 300K at 0.1 MPa [40]), and indeed $N_c$ appears to become constant within the experimental scatter. Similar behavior is approached for salol. Thus, the high temperature behavior of the dynamic correlation length scale is physically reasonable, substantiating the method of analysis.

In summary, we find that the spatial extent of the dynamic correlations for four liquids is uniquely defined by $\tau_\alpha$, independently of $T$, $P$, or $V$. This property thus mimics that of the relaxation dispersion, the $TV^\gamma$ scaling, and the onset temperatures for both non-Arrhenius behavior and the dynamic crossover, reaffirming that it is the timescale of the dynamics that unifies many prototypical properties of supercooled liquids. And while models of the glass transition generally emphasize and make predictions for $\tau_\alpha$, theories need to account for the constancy of $N_c$, $TV^\gamma$, and $\beta$ at fixed $\tau_\alpha$, since these quantities do not obviously vary the same as $\tau_\alpha$ with density, temperature, entropy, etc.

**Acknowledgements**

This work was supported by the Office of Naval Research. DF acknowledges the National Research Council for a post-doctoral fellowship.

**References**

[1] Roland C.M.; Casalini R.; Paluch M. *Chem. Phys. Lett.* **2003**, *367,* 259.

[2] Ngai K.L.; Casalini R.; Capaccioli S.; Paluch M.; Roland C.M. *J. Phys. Chem. B* **2005**, *109*, 17356.

[3] Casalini R.; Roland C.M. *Phys. Rev. E* **2004**, *69*, 062501.

[4] Alba-Simionesco C.; Cailliaux A.; Alegria A.; Tarjus G. *Europhys. Lett.* **2004**, *68*, 58.

[5] Dreyfus C.; Le Grand A.; Gapinski J.; Steffen W.; Patkowski A. *Eur. J. Phys.* **2004**, *42*, 309.

[6] Cavagna A. *Phys. Rep.* **2009**, *476*, 51.

[7] Stickel F.; Fischer, E.W., Richert R. *J. Chem. Phys.* **1996**, *104*, 2043.

[8] Angell C.A.; Ngai K.L.; McKenna G.B.; McMillan P.F.; Martin S.W. *J. Appl. Phys.* **2000**, *88*, 3113

[9] Casalini R.; Paluch M.; Roland C.M. *J. Chem. Phys.* **2003**, *118*, 5701.

[10] Casalini R.; Roland C.M. *Phys. Rev. Lett.* **2004**, *92*, 245702,

[11] Roland C.M. *Soft Matter* **2008**, *4*, 2316.

[12] Adam G.; Gibbs, J.H. *J. Chem. Phys.* **1965**, *43*, 139.

[13] Avramov I. *J. Non-Cryst. Solids* **2005**, *351*, 3163.

[14] Debenedetti, P.G.; Stillinger F.H. *Nature* **2001**, *410*, 259.

[15] Lubchenko V.; Wolynes P.G. *Ann. Rev. Phys. Chem.* **2007**, *58*, 235.

[16] Franz S.; Parisi G. *J. Phys. Condens. Matter* **2000**, *12*, 6335.

[17] Chandler D.; Garrahan J.P.; Jack R.L.; Maibaum L.; Pan A.C. *Phys. Rev. E* **2006**, *74*, 051501.

[18] Toninelli C.; Wyart M.; Berthier L.; Biroli G.; Bouchaud J.-P. *Phys. Rev. E* **2005**, *71*, 041505.

[19] Capaccioli S.; Ruocco G.; Zamponi F. *J. Phys. Chem. B* **2008**, *112*, 10652.

[20] Ediger M.D. *Ann. Rev. Phys. Chem.* **2000**, *51*, 99.

[21] Sillescu H. *J. Non-Cryst. Solids* **1999**, *243*, 81.

[22] Bohmer R. *Curr. Opin. Solid State Mat. Sci.* **1998**, *3*, 378.

[23] Chandler D.; Garrahan J.P.; Jack R.L.; Maibaum L.; Pan A.C. *Phys. Rev. E* **2006**, *74*, 051501.

[24] Berthier L.; Biroli G.; Bouchaud J.-P.; Cipelletti L. El Masri D.; L'Hote D.; Ladieu F.; Pierno M. *Science* **2005**, *310*, 1797.

[25] Franz S.; Donati C.; Parisi G.; Glotzer S.C. *Philos. Mag. B* **1999**, *79*, 1827.

[26] Berthier L.; Chandler D.; Garrahan J.P.; *Europhys. Lett.* **2005**, *69*, 320.

[27] Donati C.; Franz S.; Glotzer S.C.; Parisi G., *J. Non-Cryst. Solids* **2002**, *307-310*, 215.

[28] Dalle-Ferrier C.; Thibierge C.; Alba-Simionesco C.; Berthier L.; Biroli G.; Bouchaud J.P.; Ladieu F.; L'Hote D.; Tarjus G. *Phys. Rev. E* **2007**, *76*, 041510.

[29] Berthier L.; Biroli G.; Bouchaud J.P.; Kob W.; Miyazaki K.; Reichman D. *J. Chem. Phys.* **2007**, *126*, 184504.

[30] Ferry J.D., *Viscoelastic Properties of Polymers*; 3rd edition, Wiley: New York, 1980.

[31] Takahara S.; Ishikawa M; Yamamuro O.; Matsuo T. *J. Phys. Chem. B* **1999**, *103* 792; *103*, 3288.

[32] Hanaya M.; Hikima T.; Hatase M; Oguni M.; *J. Chem. Thermodynamics* **2002**, *34*, 1173.

[33] Moynihan C. T.; Angell C.A. *J. Non-Cryst. Solids* **2000**, *274*, 131-138.

[34] Roland C.M.; Casalini R. *J. Therm. Anal. Calorim.* **2006**, *83*, 87.

[35] Comez L.; Corezzi S.; Fioretto D.; Kriegs H.; Best A.; Steffen W. *Phys. Rev. E* **2004**, *70*, 011504.

[36] Pawlus S.; Casalini R.; Roland C.M.; Paluch M.; Rzoska S.J.; Ziolo J. *Phys. Rev. E* **2004**, *70*, 061501.

[37] Roland C.M.; Casalini R. *J. Chem. Phys.* **2005**, *122*, 134505.

[38] Roland C.M.; Capaccioli S.; Lucchesi M.; Casalini R. *J. Chem. Phys.* **2004**, *120*, 10640.

[39] Casalini R.; Paluch M.; Roland C.M. *J. Phys. Chem. A* **2003**, *107*, 2369.

[40] Stickel F.; Fischer E.W.H.; Richert R. *J. Chem. Phys.* **1995**, *102*, 6251; **1996**, *104*, 2043.

[41] Hecksher T.; Nielsen A.; Olsen N.B.; Dyre J.C. *Nature Physics* **2008**, *4*, 737.

[42] Coslovich D.; Roland C.M. *J. Chem. Phys.*, submitted [arXiv:0908.2396].

[43] Boyer R.F.; Spencer R.S. *J. Appl. Phys.* **1944**, *15*, 398.

[44] Casalini R.; Roland C.M. *Phys. Rev. B* **2005**, *71*, 014210.

[45] Casalini R.; Ngai K.L.; Roland C.M. *Phys. Rev. B* **2003**, *68*, 014201.

[46] Dalle-Ferrier C.; Eibl S.; Pappas, C.; Alba-Simionesco, C. *J. Phys.: Cond. Matter* **2008**, *20*, 494240.

[47] Richert R. *J. Chem. Phys.* **2005**, *123*, 154502.

[48] Richert R.; Duvvuri K.; Duong L.-T. *J. Chem. Phys.* **2003**, *118*, 1828.

[49] Zhu X.R.; Wang, C.H. *J. Chem. Phys.* **1986**, *84*, 6086.

[50] Zemke K.; Schmidt-Rohr K.; Magill J.H.; Sillescu H.; Spiess H.W. *Mol. Phys.* **1993**, *80*, 1317.

[51] Rajian J.R.; Huang W.R.; Richert R.; Quitevis E.L. *J. Chem. Phys.* **2006**, *124*, 014510.

FIGURE CAPTIONS

**Figure 1.** Number of molecules dynamically correlated over a time period $t \sim \tau_\alpha$ for four liquids at 0.1 MPa (hollow symbols) and at the highest pressure of the range studied (filled symbols): 500 MPa (PC), 100 (salol), 28.8 MPa (OTP-OPP), and 200 (PCB). The temperature and pressure ranges were determined by the availability of experimental data.

**Figure 2.** Number of correlated molecules as a function of the log of the relaxation time; At any $\tau_\alpha$, values of $N_c$ for different pressures are equivalent within the experimental uncertainty. Pressures (in MPa) are 0.1 (open squares), 250 (crossed squares), and 500 (filled squares) for PC; 0.1 (open circles), 100 (crossed circles), and 200 (filled circles) for PCB; 0.1 (open inverted triangles), 200 (crossed inverted triangles), and 400 (filled inverted triangles) for salol; 0.1 (open triangles) and 28.8 (filled triangles) for OTP-OPP.

**Figure 3.** (bottom) Number of correlated molecules plotted in Arrhenius form. PCB and salol show a change in behavior at the dynamic crossover (indicated by arrows [45]), although there is no obvious change for PC at $T_B$. At the onset of Arrhenius behavior ($\beta(T_A) = 1$), $N_c$ for PC and salol approach a limiting low value. (top) Variation of Kohlrausch exponent with temperature; smooth lines were used to interpolate where gaps existed in the experimental values of $\beta$.

Figure 1.

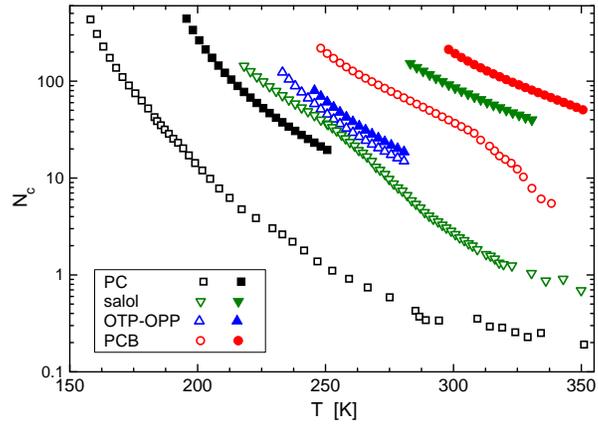

Figure 2.

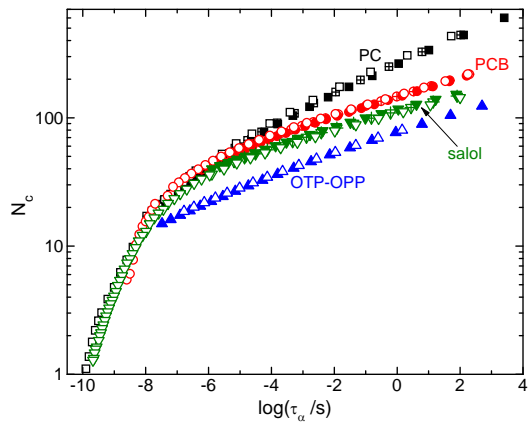

Figure 3.

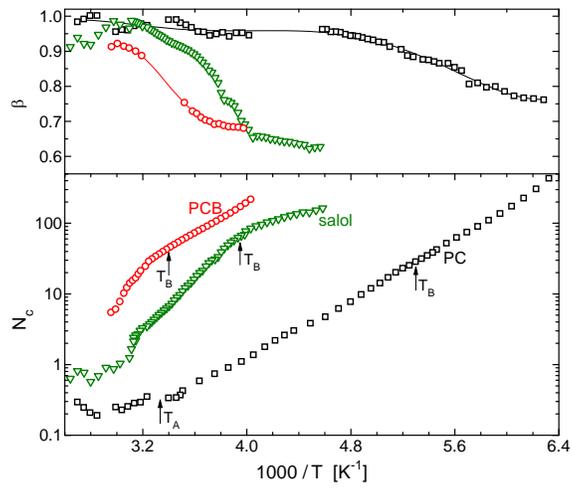

11